\documentclass[aps,prc,twocolumn,superscriptaddress,nofootinbib]{revtex4-2}
\usepackage{graphicx}
\usepackage{appendix}
\usepackage{amsmath}
\usepackage{comment}
\usepackage{float}
\usepackage{inputenc}
\usepackage{fontenc}
\usepackage[bookmarks,bookmarksnumbered]{hyperref}
\hypersetup{colorlinks = true,
            linkcolor = blue,
            anchorcolor =red,
            citecolor = blue,
            pdfauthor=author}

\usepackage[dvipsnames]{xcolor}
\usepackage{lineno}

\graphicspath{{figures/}}
\usepackage{color}

\newcommand{\bth}{\boldsymbol{\theta}}

\newcommand{\Npc}{N_{\mathrm{pc}}}

\begin{document}
\title{
Examination of nucleon distribution with Bayesian imaging for isobar collisions}
\author{Yi-Lin Cheng}
\affiliation{Frankfurt Institute for Advanced Studies (FIAS), D-60438 Frankfurt am Main, Germany.}
\affiliation{Shanghai Institute of Applied Physics, Chinese Academy of Sciences, Shanghai 201800, China.}
\affiliation{Key Laboratory of Nuclear Physics and Ion-beam Application (MOE),
Institute of Modern Physics, Fudan University, Shanghai 200433, China.}
\affiliation{University of Chinese Academy of Sciences, Beijing 100049, China.}

\author{Shuzhe Shi}
\email{shuzhe-shi@tsinghua.edu.cn}
\affiliation{Department of Physics, Tsinghua University, Beijing 100084, China.} 
\affiliation{Center for Nuclear Theory, Department of Physics and Astronomy, Stony Brook University, Stony Brook, New York 11794-3800, USA.}

\author{Yu-Gang Ma}
\affiliation{Key Laboratory of Nuclear Physics and Ion-beam Application (MOE),
Institute of Modern Physics, Fudan University, Shanghai 200433, China.}

\author{Horst St\"ocker}
\affiliation{Frankfurt Institute for Advanced Studies (FIAS), D-60438 Frankfurt am Main, Germany.}
\affiliation{Institute f\"ur Theoretische Physik, Goethe Universit\"at, D-60438 Frankfurt am Main, Germany.}
\affiliation{GSI Helmholtzzentrum f\"ur Schwerionenforschung GmbH, D-64291 Darmstadt, Germany.}

\author{Kai Zhou}
\email{zhou@fias.uni-frankfurt.de}
\affiliation{Frankfurt Institute for Advanced Studies (FIAS), D-60438 Frankfurt am Main, Germany.}

\begin{abstract}
Relativistic collision of isobaric systems is found to be valuable in differentiating the nucleon distributions for nuclei with the same mass number. In recent contrast experiment of $^{96}_{44}\text{Ru}+^{96}_{44}\text{Ru}$ versus $^{96}_{40}\text{Zr}+^{96}_{40}\text{Zr}$ collisions at $\sqrt{s_\text{NN}} = 200~\text{GeV}$, the ratios of multiplicity distribution, elliptic flow, triangular flow, and radial flow are precisely measured and found to be significantly different from unity, indicating the difference in the shapes of the isobar pair. In this work, we investigate the feasibility of nuclear structure reconstruction from heavy-ion collision observables.
We perform Bayesian Inference with employing the Monte-Carlo Glauber model as an estimator of the mapping from nuclear structure to the final state observables and to provide the mock data for reconstruction. By varying combination of observables included in the mock data, we find it plausible to infer Woods--Saxon parameters from the observables. We also observe that single-system multiplicity distribution for the isobar system, rather than their ratio, is crucial to simultaneously determine the nuclear structure for the isobar system.
\end{abstract}
\maketitle

\section{Introduction}
Strongly coupled QCD matter can be studied by relativistic heavy ion experiments, which have been carried out for decades at the Relativistic Heavy Ion Collider (RHIC) of Brookhaven National Laboratory~\cite{STAR:2005gfr} and Large Hadron Collider (LHC) of CERN~\cite{Muller:2012zq} facilities.
An important scientific breakthrough from both experimental and theoretical efforts, among others, is the revealed extremely hot quark-gluon plasma (QGP) in these collisions (see e.g.~\cite{Chen:2018tnh, Busza:2018rrf}). Large-scale collective motion is formed at the partonic level for a short period of time and converted to the final state observables, such as elliptic, triangular, and radial flows.

The QGP initial state, which is dictated by the nuclear structure with its intrinsic deformation characterized by spherical, ellipsoidal, octuple and hexadecapole modes~\cite{RevModPhys.83.1467, Togashi:2016yzs, Heyde:2016sop, Frauendorf:2017ryj}, can be featured by its size and eccentricities $\varepsilon_{n}$ and further transferred into radial and azimuthal anisotropic flow of final state hadrons~\cite{Ollitrault:1992bk, Voloshin:1994mz, Qiu:2011iv, Heinz:2013wva, Filip:2009zz, Alver:2010gr, Carzon:2020xwp}. Consequently, the observables measured in the final stage of heavy-ion collisions, such as elliptic flow $v_{2}$, triangular flow $v_{3}$ and charged particle multiplicity $N_{ch}$, have a strong relationship with the initial state characterized by second and third order anisotropy, $\varepsilon_{2}$ and $\varepsilon_{3}$~\cite{Zhang:2017xda}, and the total energy,
and thus also hold an imprint of nuclear structure especially the deformation~\cite{Rosenhauer:1986tn, Shuryak:1999by}.

Recently a high statistics contrast heavy-ion collision, which collides $_{44}^{96}$Ru+$_{44}^{96}$Ru and $_{40}^{96}$Zr+$_{40}^{96}$Zr with beam energy $\sqrt{s_\text{NN}} = 200~\text{GeV}$, is performed by the STAR Collaboration at RHIC~\cite{STAR:2021mii}\footnote{For the rest of the paper, we will refer to the $_{44}^{96}$Ru and $_{40}^{96}$Zr nucleus as Ru and Zr, separately.}. It was originally designed to search for the chiral magnetic effect (CME)~\cite{Kharzeev:2004ey, Kharzeev:2007jp, Fukushima:2008xe,STAR:2019bjg}, under the presumption that the different electric number would induce a sizable difference in the CME signal while the same baryon number would lead to the same non-CME background. 
Unexpectedly, significant differences are observed in the bulk properties~\cite{STAR:2021mii}, which prevents one to make a conclusive statement on the existence of CME in heavy-ion collisions and calls for more efforts in a better quantification of the background~\cite{Feng:2021pgf, Kharzeev:2022hqz}. 
However, it brings a new opportunity to study the nucleon distribution in relativistic heavy-ion collisions~\cite{Giacalone:2019pca, Giacalone:2020awm, Jia:2021oyt, Jia:2021tzt, Zhang:2021kxj, Jia:2022iji, Giacalone:2021udy, Li:2019kkh, Xu:2021uar, Xu:2021qjw, Xu:2021vpn, Zhao:2022uhl, Bally:2022vgo, Nijs:2021kvn}.

The nucleon distribution within a nucleus is often described by a deformed Woods--Saxon distribution, 
\begin{equation}
    \rho(r, \theta, \phi)=\frac{\rho_{0}}{1+\exp{[r-R(\theta, \phi)]/a}}
    \label{eq:woodssaxon}
\end{equation}
where
\begin{equation}
    R(\theta, \phi)=R_{0}(1+\beta_{2}Y^{0}_{2}+\beta_{3}Y^{0}_{3}+\beta_{4}Y^{0}_{4}), 
\end{equation}
with $R_{0}$ is the radius, $a$ the skin depth, and $\beta_2$, $\beta_3$, and $\beta_4$ respectively the quadruple, octupole, and hexadecapole momentums~\cite{Shou:2014eya, Jia:2021tzt}. $Y_{l}^{m}(\theta,\phi)$ are the spherical harmonics, describing the angular dependence of the nuclear radius. Both $a$ and $R_{0}$ have significant impacts on the initial overlap area in the collision systems~\cite{Zhang:2022fou, Xu:2021uar}.
In ultra-central collisions, the ellipticity $\varepsilon_{2}$ and triangularity $\varepsilon_{3}$ are correlated to the quadrupole $\beta_{2}$ and octupole $\beta_{3}$ deformations, respectively~\cite{Kuhlman:2005ts}. Meanwhile, the mean energy density in the overlap of the heavy ion collision, which measures the pressure gradient, is expected to be correlated with mean-$p_T$~\cite{Jia:2021qyu}. Therefore, the nuclear structure can be reflected in the final state observables of heavy-ion collisions~\cite{Ma:2022dbh}.

It is proposed to study the ratio of observables in the isobar systems~\cite{Zhang:2021kxj}, in order to reduce the medium expansion's systematic uncertainties of, e.g., transport properties. While a set of parameters has been conjectured which describes the ratios data roughly, systematic/principled improvement and the uniqueness of the parameter set remains unclear. In this work, we aim to answer two questions: 
\begin{itemize}
    \item[1)] is it possible to infer initial state nucleon distribution thus the nuclear structure information from final state observables in single collision system?
    \item[2)] is it possible to simultaneously reconstruct the nuclear structures of isobar systems from a contrast isobar collision experiment? 
\end{itemize}

In order to answer these questions, we take the Monte Carlo Glauber model as the estimator to provide initial-to-final mapping in heavy ion collisions and also generate mock data. Then we perform Bayesian Inferences of the Woods--Saxon parameters based on different combinations of observables.
The inference framework design will be discussed in Sec.~\ref{sec:framework} whereas results can be found in Sec.~\ref{sec:result}. Finally we summarize in Sec.~\ref{sec:summary}.

\section{Bayesian Inference of Nuclear Structure}~\label{sec:framework}
In this work, we investigate the feasibility of constraining nuclear structure information via Bayesian inference from the accessible physical observables in heavy-ion collisions, e.g., multiplicity distribution, anisotropic flows, and radial flow (mean-$p_T$) related information. Later in subsection~\ref{sec:2.mcglb} we will detail the to-be-used observables in this analysis, which let us denote collectively as $\mathcal{D}$ for sake of clarity. About the target, to be specific we aim at inferring the nuclear structure parameters (collectively denoted as $\boldsymbol{\theta}$) in the Woods-Saxon distribution for single\footnote{We note that such a procedure takes the Ansatz that the observables of interest depend only on the nucleon distribution, but not other properties of the nuclear such as electric charge or spin.}. or combined isobaric collision system analysis, including the radius $R$, surface diffuseness $a$, deformation quadrupole $\beta_2$ and octupole $\beta_3$. This constructs an inverse problem actually, which might also be tackled alternatively by deep learning based approaches developed e.g. in ~\cite{Pang:2019aqb,Shi:2021qri,Soma:2022qnv,Wang:2021jou,Shi:2022yqw}. We focus on the Bayesian inference for this task in the present work and leave other methods solving it for future work.

Bayesian inference is a statistical approach to update knowledge about the targeted physical parameters inside a computational physical model based upon evidence of observed data. It has been applied in heavy-ion physics for religious determination of transport parameters~\cite{Bernhard:2016tnd, Bernhard:2019bmu, JETSCAPE:2020shq, JETSCAPE:2020mzn, JETSCAPE:2021ehl, Nijs:2020roc, Nijs:2020ors, Parkkila:2021tqq} and EoS~\cite{Pratt:2015zsa, OmanaKuttan:2022aml}. According to the Bayes theorem, the posterior distribution of the parameters $\boldsymbol{\theta}$ given the observed data of $\mathcal{D}$ can be expressed as, 
\begin{align}
    p(\boldsymbol{\theta}|\mathcal{D})\propto p(\mathcal{D}|\boldsymbol{\theta})p(\boldsymbol{\theta}),
\end{align}
where $p(\boldsymbol{\theta})$ is the prior distribution encoding our initial knowledge of the parameters and $p(\mathcal{D}|\boldsymbol{\theta})$ is the likelihood distribution representing how good any chosen parameters are in describing the observed data. In such a way with the Bayesian statistics perspective, the obtained posterior $p(\boldsymbol{\theta}|\mathcal{D})$ codifies the updated knowledge on the parameters $\boldsymbol{\theta}$ after confronting the observation of data $\mathcal{D}$, from which the nuclear structure parameters can then be sampled. 

We follow state-of-the-art general procedures of Bayesian inference for estimating parameters in computationally intensive models~\cite{doi:10.1198/016214507000000888, Higdon:2014tva}, including briefly, model evaluation on some representative ``design points'' in the parameter space, emulator training for mimicking the model simulation in a more efficient manner, and MCMC sampling to construct and exploit the posterior distribution of the parameters. This Bayesian approach has been applied in a number of studies for heavy-ion collisions for constraining especially the dynamical parameters like shear viscosity.
In the rest of the section, we will explain the detailed procedure of each step.

\subsection{Data Preparation using Monte Carlo Glauber}\label{sec:2.mcglb}
As an exploratory first step Bayesian study on imaging nuclear structure from relativistic heavy-ion collision measurements, the present work focuses on the possibility of inferring the nuclear structure parameters from heavy-ion collision initial state information with well-understood correspondence to the final state observables. To this end, we employ the Monte Carlo Glauber(MC-Glauber) as an estimator for the heavy ion collision initial state observables. It is worth noting that for isobaric collisions the ratio of many observables like anisotropic flows are argued to be insensitive to the medium evolution and thus highlighting the sole relevance of initial state information of the collisions.

The MC-Glauber model~\cite{Broniowski:2007nz, Alver:2008aq} is a widely used model to generate event-wise fluctuating initial conditions for heavy-ion collisions. It starts from sampling the nucleon configuration according to the Woods--Saxon distribution~\eqref{eq:woodssaxon}, determines the binary collision pairs according to their transverse separation, and randomly assigns energy deposition, according to a gamma distribution, for each binary collision pair and for each participant nucleons. Eventually, one obtains the energy density distribution in the transverse plane, $e(x,y)$, which can be treated as the initial condition of the hydrodynamic evolution. In this work, the parameters in the MC-Glauber model are taken from Ref.~\cite{Shen:2014vra}.

For the purpose of the present study, we do not feed the MC-Glauber initial conditions into hydro or transport evolution and compute the final state observables. Instead, we use the linear response approximation to estimate observables from $e(x,y)$ (see e.g.~\cite{Jia:2021qyu}). In other words, we employ MC-Glauber to provide a mapping from the nuclear structure to final state observables and investigate the feasibility of reconstructing the former from the latter using Bayesian Inference. Such a procedure is reliable when we aim to know whether final state observables can be used to infer the nuclear structure (and if possible, what observables are needed) without aiming to unbiasedly extract the nuclear structure from the current data. 
A study for the latter purpose requires dynamical evolution models, such as AMPT~\cite{Lin:2021mdn} or hydrodynamics~\cite{Shen:2020mgh}, and will be reported in our follow-up paper.

Final state observables can be estimated from the initial state energy distribution through relations listed as follows (see e.g.~\cite{Jia:2021qyu}). 
First, the charged multiplicity can be estimated from the total energy, 
\begin{align}
    N_\mathrm{ch} \propto E \equiv \int e(r,\phi) \,r\,\mathrm{d}r\,\mathrm{d}\phi,
\end{align}
which is natural due to energy conservation. 
Second, the two-particle elliptic and triangular flows can be approximated by the initial anisotropies, $v_{n}\{2\} \propto \sqrt{\langle\varepsilon_{n}^2\rangle}$, where  $\langle\cdot\rangle$ refers to the average over events within the same multiplicity bin, $n=2$ or $3$, and
\begin{align}
    \varepsilon_{n} \equiv \frac{|\int e(r,\phi) r^n e^{i\,n\phi} r\,\mathrm{d}r\,\mathrm{d}\phi|}{\int e(r,\phi) r^n r\,\mathrm{d}r\,\mathrm{d}\phi}\,.
\end{align}
Finally, one may estimate the radial flow $\langle p_T\rangle$ from a monotonic function of the energy density, 
\begin{align}
     d_\perp \equiv \sqrt{E/S_\perp},
\end{align}
where $S_\perp \equiv \pi\sqrt{\frac{\int e(x,y)x^2 \mathrm{d}x\,\mathrm{d}y}{E}\frac{\int e(x,y)y^2 \mathrm{d}x\,\mathrm{d}y}{E}}$ is the energy weighted transverse area. Therefore, we respectively take $E$, $\varepsilon_2$, $\varepsilon_3$, and $d_\perp$ as the estimator of $N_\mathrm{ch}$, $v_2\{2\}$, $v_3\{2\}$, and $\langle p_T \rangle$, and may interchangeably use two sets of notations in the succeeding text. 

To prepare the training data, we sampled 1001 ``design points'' from the four-dimension parameter space spanned over $R\in[4.9,5.2]~\mathrm{fm}$, $a\in[0.3,0.6]~\mathrm{fm}$, $\beta_2\in[0,0.4]$, and $\beta_3\in[0,0.4]$, according to Latin hypercube sampling which ensures uniform space-filling. For each parameter set, we simulate $10^6$ events with unequal weights that enhance the probability of ultra-central collision events. That is, rather than the desired probability distribution\footnote{The desired probability distribution is derived from the Jacobian, $b\,\mathrm{d}b$.}, $P(b) \propto b$, we sample the impact parameter ($b$) of an event according to $\bar{P}(b) \propto e^{-b/6}$ and then each event will be assigned with a weight $w=b\, e^{b/6}$. After the simulation, we bin all events according to $N_\mathrm{ch}$ with 40 intervals: $1 \leq N_\mathrm{ch} \leq 10$, $11 \leq N_\mathrm{ch} \leq 20$, $\cdots$, $391 \leq N_\mathrm{ch} \leq 400$. In each bin, we calculate the event counts and mean values of elliptic, triangular, and radial flows and their corresponding uncertainties, with weights correctly considered. For later convenience, we represent the observables from simulation by an array with 160 elements, which is $\boldsymbol{O}\equiv\{P_{1 - 10}, \cdots, P_{391 - 400},
v_{2,{1 - 10}},\cdots,$ $v_{3,{1 - 10}},\cdots, \langle p_T\rangle_{{1 - 10}},\cdots,\langle p_T\rangle_{{391 - 400}}\}$. We further denote the simulation results as $O_{a,i}$, with the index $i$($a$) labeling the simulation parameter(observable). In other words, $O_{a,i}$ represents the $a$-th observable for simulation take the parameter set $\boldsymbol{\theta}_i\equiv\{R_i, a_i, \beta_{2,i}, \beta_{3,i}\}$. We adopt $d=1001$ and $m=160$ to represent the number of events sampled in the parameter space and dimensions of observables, respectively.

\subsection{Postprocess of Model Outputs}
With the simulation results, $O_{a,i} = O_a(\boldsymbol{\theta}_i)$, which forms the training data set, our next task is to get a fast surrogate by decoding out an approximation function $O_a(\boldsymbol{\theta})$ out of it. We note that they are not independent functions as one would expect correlations exist between them. For instance, the continuity of observables between neighboring multiplicity bins is an obvious example of the correlation. 
Therefore, we first transform the model outputs into a smaller number of uncorrelated variables using the principal component analysis (PCA) method, then we treat each new principal component as independent ``observables''. The PCA method performs linear combinations of observables and ranks them according to their sensitivity to the parameter variation. 

To find out the features that are most sensitive to the change in parameter and avoid bias from the overall magnitude of different observables, we standardize the training data set by subtracting their mean and also dividing their standard deviation over all parameters,
\begin{align}
    \widetilde{O}_{a,i} \equiv \frac{{O}_{a,i}- \mu_a}{ \sigma_a},
\end{align}
where
\begin{align}
    \mu_a = d^{-1}\sum_{i=1}^{d} {O}_{a,i}, 
    \qquad
    \sigma_a^2 = d^{-1}\sum_{i=1}^{d} ({O}_{a,i} - \mu_a)^2. 
\end{align} 
Then we compute the covariance matrix
\begin{align}
\mathrm{\widetilde C}_{ab} 
=\;& 
     d^{-1} \sum_{i=1}^{d} \widetilde{O}_{a, i}\, \widetilde{O}_{b,i}
\end{align}
and diagonalize it as
\begin{align}
    \sum_{a,b=1}^{m} V_{af} \widetilde{\mathrm{C}}_{ab} V_{bf'} = \lambda_{f} \delta_{ff'}.
\end{align}
Here, $\boldsymbol{V}$ is an orthogonal matrix,
and eigenvalues are ranked according to the importance $\lambda_{1} \geq \lambda_{2} \geq \lambda_{3} \geq \cdots$, such that the first principal component has the maximum possible variance, which can explain as much variance of $\mathrm{\widetilde C}$ as possible, and the second component has maximal variance while being orthogonal to the first, and so forth. 
Then the principal components are constructed as the eigenvectors of the covariance matrix, 
\begin{align}
    \mathrm{PC}_{f,i} = \sum_{a=1}^{m} V_{af} \widetilde{O}_{a,i}, 
\end{align}
which obeys the condition that
\begin{align}
    \sum_{i=1}^{d} \mathrm{PC}_{f,i} = 0, 
    \qquad
    \sum_{i=1}^{d} \mathrm{PC}_{fi} \mathrm{PC}_{f'i} = d\, \lambda_f\, \delta_{ff'}. 
\end{align}

Therefore, when varying the parameters, different $\mathrm{PC}_{f}$'s are independent of each other. They will be fed into the Gaussian Process emulator to learn the function mappings from input nuclear structure parameter to PC's [$\mathrm{PC}_{f}(\bth)$], with details will be discussed in the succeeding subsection. Once the approximation function is obtained,
observables can be reconstructed from the inverse transformation with uncertainties as well,
\begin{align}
\begin{split}
O_{a}(\bth)
=\;&
    \mu_a + \sigma_a\, \sum_{f=1}^{\Npc} V_{af} \mathrm{PC}_{f}(\bth)\, , 
\end{split}\\
\begin{split}
\delta O_{a}(\bth)
=\;&
\sigma_a\ \sqrt{\sum_{f=1}^{\Npc} V_{af}^{2} \delta\mathrm{PC}_{f}^{2}(\bth)}\,.
\end{split}
\end{align}

\begin{figure}[!hbtp]
    \centering
    \includegraphics[width=0.45\textwidth]{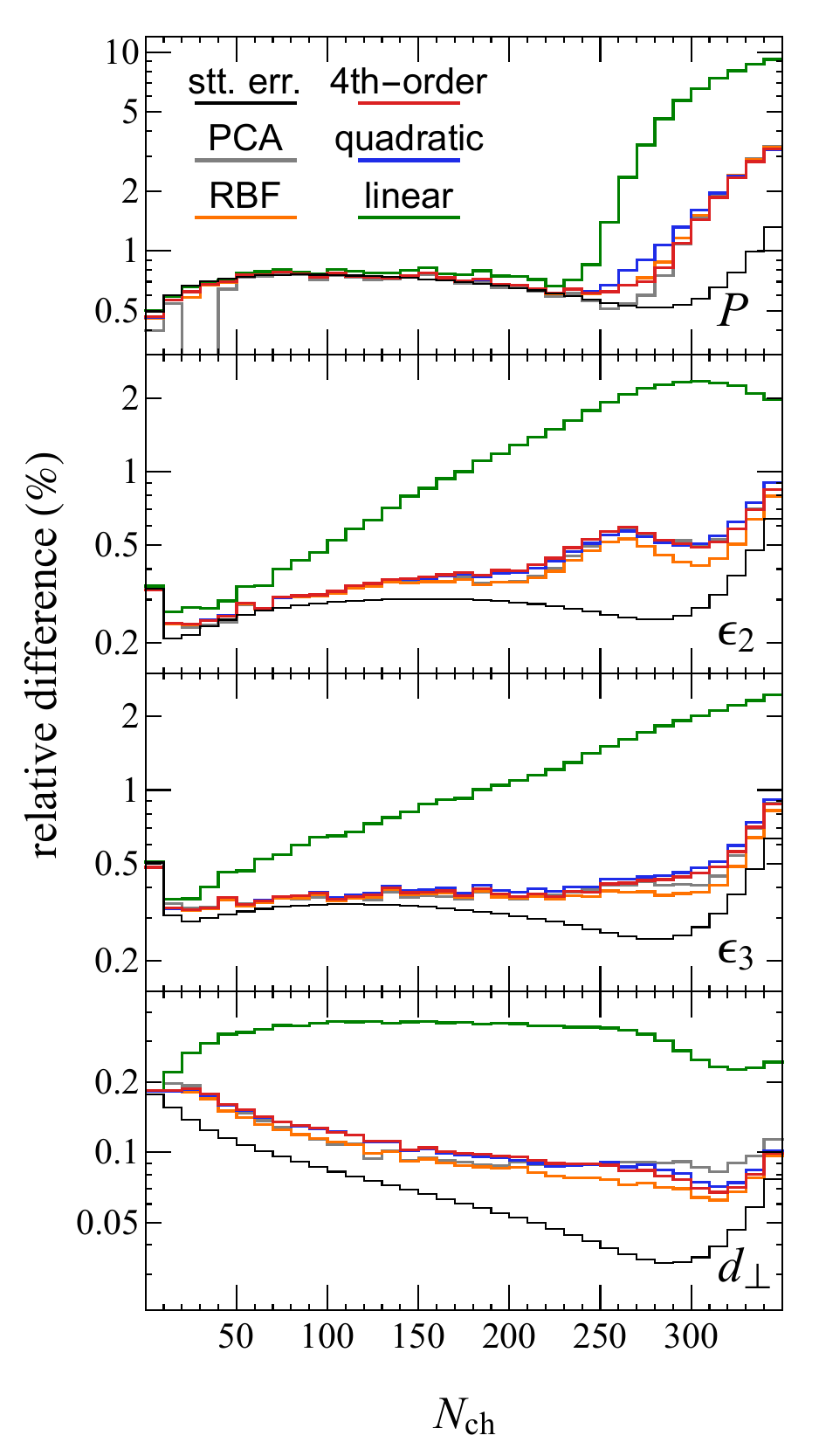}
    \caption{Comparison of relative difference~\protect{\eqref{eq:diff_rel}} between the ground truth and predicted values using Gaussian Processor with linear(green), quadratic(blue), $4^\text{th}$-order(green) and RBF(orange) kernels. 
    As references, gray curves represent the differences due to the PCA truncation. Statistical errors over the mean value in the MC-Glauber modeling are also presented as black curves.
    \label{fig:difference}}
\end{figure}

In our study, despite there being $160$ original observables, we find that the first and the second PC explain over $55\%$ and $24\%$ of the original variance respectively, and the number increases to $96\%$ when the first four PCs are considered. We adopt the first ten PCs (i.e., make the truncation with $\Npc = 10$) which cover $98\%$ of the total variance. It shall be worth noting that such truncation can lead to distortion of the training data, i.e., 
\begin{align}
\begin{split}
    \Big|\mu_a + \sigma_a\, \sum_{f=1}^{\Npc} V_{af} \mathrm{PC}_{f,i} - O_{a,i} \Big| \geq 0,
\end{split}
\end{align}
with equality is taken when $\Npc=160$. We examine that taking $\Npc=10$ would lead to differences that is comparable to the statistical errors in the MC-Glauber simulation, and hence the distortion is negligible. See gray curves in Fig.~\ref{fig:difference}.

\subsection{Gaussian Process Emulator} \label{GP}
An empirical linear relation between observables in heavy-ion collisions and nuclear structure parameters has been found in Ref.~\cite{Jia:2021oyt} for small variation of parameters,
\begin{equation}
    O\approx b_{0}+b_{1}\beta_{2}^{2}+b_{2}\beta_{3}^{2}+b_{3}(R_{0}-R_{0, ref})+b_{4}(a-a_{ref}),
    \label{eq:empirical}
\end{equation}
where $b_{0}$ represents the value for spherical nuclei at some reference radius and diffuseness, and $b_{1}$, $b_{2}$, $b_{3}$, $b_{4}$ are centrality dependent response coefficients that encode the final state dynamics. While Eq.~\eqref{eq:empirical} provides a valuable hint to approximate the function $\mathrm{PC}_{f}(\bth)$, one shall be careful in its applicability in a wide range of parameter sets. In this work, we adopt the Gaussian Process(GP) regression to construct one emulator to fast evaluate the observables, where we also try different kernels to systematically study observable responses. 

Rather than calculating the probability distribution of parameters of a specific function, GP regression calculates the probability distribution over all admissible functions within chosen covariance behavior to fit the data.
A Gaussian process is defined as a collection of random variables, any finite number of which have a joint Gaussian distribution. 
With the training parameter sets $\{\bth_1, \bth_2, \cdots, \bth_d\}$ and their corresponding PC\footnote{Since this will be performed independently for each PC, we have omitted the subscript $f$ which labels them.} of observables $\{\mathrm{PC}(\bth_1), \mathrm{PC}(\bth_2),\cdots,\mathrm{PC}(\bth_d)\}$, the function value at a particular parameter point, $\mathrm{PC}(\bth)$, follows the Gaussian distribution,
\begin{align}
     \mathrm{PC}(\bth) \sim \mathcal{N} (\overline{\mathrm{PC}}(\bth), \sigma^2(\bth)),
     \label{eq:GP_distribution}
\end{align}
where the mean function $\overline{\mathrm{PC}}(\bth)$ and the variance $\sigma^2(\bth)$ are determined by the training date set and the convariance function, also called kernel, $k(\bth, \bth')$. The kernel encodes our assumptions about the function $\mathrm{PC}(\bth)$ to be learned. 
We introduce a matrix $\boldsymbol{K}$, in which the $(i,j)$-th element is given by $K_{i,j}=k(\bth_i, \bth_j)$. Then the mean and variance are given by~\cite{Rasmussen:2006gp}
\begin{align}
\overline{\mathrm{PC}}(\bth) =\;& \sum_{i,j=1}^{d} k(\bth,\bth_i) (K^{-1})_{i,j} \mathrm{PC}(\bth_j)\,,\\
\sigma^2(\bth) =\;& k(\bth,\bth) - \sum_{i,j=1}^{d} k(\bth,\bth_i) (K^{-1})_{i,j} k(\bth_j,\bth)\,.
\end{align}

Usually, before drawing functions from a GP, we must specify the covariance function. A standard choice for kernel is the radial basis function kernel (RBF), which is also known as the squared exponential (SE) covariance function,
\begin{align}
     k_\mathrm{RBF}(\bth_{i},\bth_{j})
     =\exp\Big(-\frac{ d(\bth_{i},\bth_{j})^2}{2l^{2}}\Big)\,,
     \label{eq:RBF}
\end{align}
where $l$ is the length scale of the kernel, which equals unity in the current work, and $d(\bth_i, \bth_j)=\sqrt{|\bth_i - \bth_j|^2}$ is the Euclidean distance. In addition, in order to study the observable's response to the change of Woods--Saxon parameters $\bth$, we consider also inhomogeneous polynomial kernels for the GP emulator, including linear, quadratic, and fourth-order regression kernels:
\begin{align}
k_\mathrm{lin}(\boldsymbol \bth_{i}, \boldsymbol \bth_{j}) 
=\;&
\sigma_{0}^{2}+\boldsymbol \bth_{i} \cdot \boldsymbol \bth_{j},
\label{eq:linear}\\
k_\mathrm{quad}(\bth_i, \bth_j) 
=\;& 
k_\mathrm{lin}^2(\boldsymbol \bth_{i}, \boldsymbol \bth_{j}),
\label{eq:qua}\\
k_\mathrm{fourth}(\bth_i, \bth_j)
=\;&
k_\mathrm{lin}^4(\boldsymbol \bth_{i}, \boldsymbol \bth_{j}),
\label{eq:four}
\end{align}
where $\sigma_{0}$ controls the inhomogenity of the kernel and is set as unity here. 
Meanwhile, a white kernel, $k_\mathrm{white}(\boldsymbol \bth_{i},\boldsymbol \bth_{j})=3\,\delta_{ij}$, is also added into the covariance function in GP emulator to account for the fluctuation noise induced by the omission of higher principle components.
So we combine the above polynomial kernel and white kernel as the covariance function for linear, quadratic, and fourth-order \textit{regression} respectively. More details of the GP regression and different kernels' influence can be found in Ref.~\cite{Rasmussen:2006gp}.

To examine the accuracy of the GP emulators, we quantify the relative uncertainty of the emulator as the square of the relative difference between the true value ($O_{a,i}^\mathrm{truth}$) and the emulator prediction ($O_{a,i}^\mathrm{pred}$), averaged over the testing ensemble,
\begin{align}
    \delta O_{a}^\text{rel} \equiv \sqrt{\frac{1}{d}\sum_{i=1}^{d} \Big(\frac{O_{a,i}^\mathrm{pred}-O_{a,i}^\mathrm{truth}}{O_{a,i}^\mathrm{truth}}\Big)^2}. 
    \label{eq:diff_rel}
\end{align}
In Fig.~\ref{fig:difference} we compare the relative differences of different observables taking linear, quadratic, and fourth-order regression kernels in GP. For comparison, we also tried the widely-adopted RBF kernel (\ref{eq:RBF}), which can interpolate the training data points smoothly. We find that quadratic and fourth-order \textit{regressions} are able to reach the accuracy of RBF \textit{intepolation}, all of which are close to the ``upper bound'' of performance limited by the PC truncation. We also note that results taking linear \textit{regression} agree decently with the simulation data, but its significantly larger difference indicates the limitation of applying~\eqref{eq:empirical} in such a wide range of parameters, and therefore higher moments are needed. The similarity between quadratic and fourth-order \textit{regression} shows the convergence of the expansion. In this work, we set the emulator of the function relation between Woods--Saxon parameters and observables in heavy-ion collisions to be the fourth-order regression kernel, which not only ensures accurate description but also has a clear physics dependence on the Woods--Saxon parameters, i.e., a higher order generalization of the empirical linear response~\eqref{eq:empirical}.

\begin{figure*}[!hbtp]
    \centering
    \includegraphics[width=0.4\textwidth]{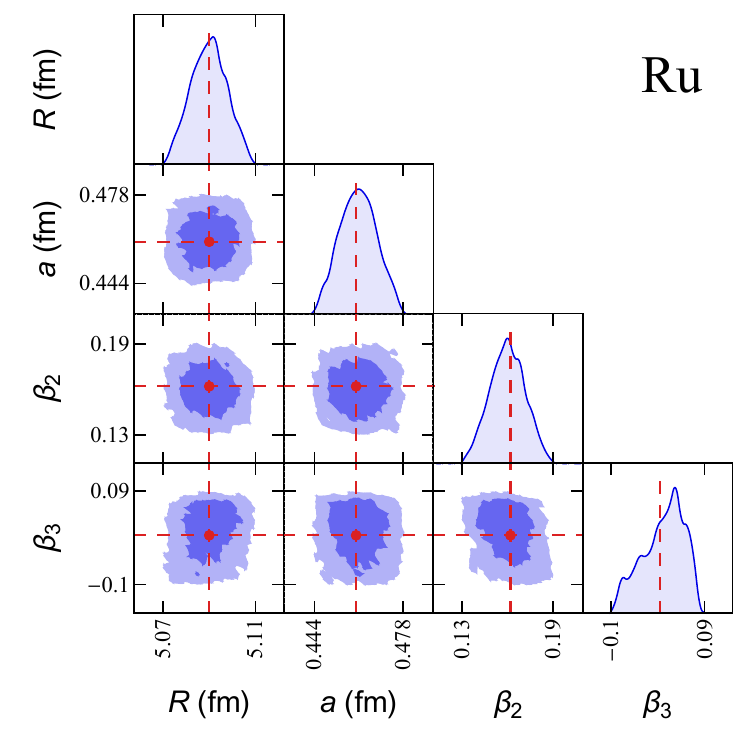}\qquad\qquad\qquad
    \includegraphics[width=0.4\textwidth]{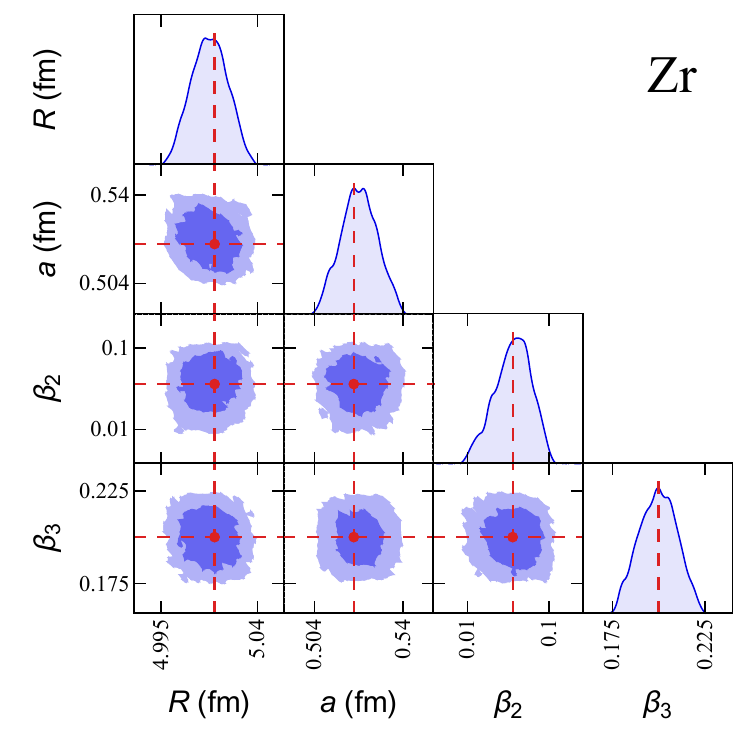}
    \caption{One- and two-dimensional marginal posterior distributions of nuclear structure parameters reconstructed from single system mock data corresponding to Ru-Ru(left) and Zr-Zr(right) collisions. In the two-dimensional posterior, darker(lighter) area correspond to the $68\%$($95\%$) C.L. regions.
    \label{fig:reconstruction_single}}
\end{figure*}

\subsection{Parameter Inference using Markov Chain Monte Carlo}
With the function relation $O_a(\boldsymbol{\theta})$ represented by the GP emulator, now we are ready to infer the likelihood distribution of parameters from observables. According to Bayes' theorem, given experimental measurements $y_a^\mathrm{exp}$ with covariance matrix $\Sigma_\mathrm{exp}^{-1}$, the Poseterior distribution of model parameter is given by 
\begin{align}
&P(\boldsymbol{\theta}|y_a^\mathrm{exp})
= N\exp\Big(-\frac{\chi^2}{2}\Big)\mathrm{Prior}(\boldsymbol{\theta})\,
\label{eq:posterior}\\
&\chi^2 = \sum_{a,b} (\Sigma_{\boldsymbol{\theta}}^{-1})_{a,b} 
    \Delta y_a(\boldsymbol{\theta}) \Delta y_b(\boldsymbol{\theta})\,,
\label{eq:chisq_mat} 
\end{align}
where $N$ is the normalization parameter, $\Delta y_a(\boldsymbol{\theta}) \equiv y_a(\boldsymbol{\theta}) - y_a^\mathrm{exp}$ is the difference between experimental observables and the corresponding model prediction. Observables, herein, can be either single system observable $O_a(\boldsymbol{\theta})$ or the ratio in isobar system.
$\Sigma_{\boldsymbol{\theta}} = \Sigma^\mathrm{exp} + \Sigma^{\mathrm{emu}}_{\boldsymbol{\theta}}$ is the covariance matrix which includes the contribution from both experimental and theoretical(emulator) uncertainties. We note that we keep only the most important PC's in the emulator, which makes it nontrivial to calculate the correlation between ratios. Details will be shown in Appendix~\ref{sec:correlation}.
$\mathrm{Prior}(\boldsymbol{\theta})$ is the prior distribution that encodes the preknowledge of the parameters. We take uniform prior distribution within the parameter range of the training data set.

In this work, we employ the Markov chain Monte Carlo(MCMC) method to compute $P(\boldsymbol{\theta}|y_a^\mathrm{exp})$. 
MCMC generates representative samples according to the posterior distribution by making a random walk in parameter space weighted by the relative posterior probability. 
Denoting the parameter set in the $n^\mathrm{th}$ iteration step as $\boldsymbol{\theta}^{(n)}$, MCMC samples the next step $\boldsymbol{\theta}^{(n+1)}$ as a random walk starting from $\boldsymbol{\theta}^{(n)}$, and decides whether to accept this update according to a probability being $p=\mathrm{min}[1, P(\boldsymbol{\theta}^{(n+1)})/P(\boldsymbol{\theta}^{(n)})]$. 
One can show that samples generated in such a way satisfy the desired posterior distribution $P(\boldsymbol{\theta}|y_a^\mathrm{exp})$. 
Markov chains have a property that the conditional probability distribution of future states depends only on the current state and not on the past states preceded the current one, which guarantees the stational distribution to be achieved.

For the practical computation, we apply the \texttt{pymc} framework and choose the No-U-Turn Sampler(NUTS), which is the extension to Hamiltonian Monte Carlo method~\cite{1507.08050, 1111.4246}. We take 1000 samples from each Markov chain with iteration number of adaptive phase (i.e., number of update iterations between two taken samples in a Markov Chain) being 1500 to ensure decorrelation. With a sufficient amount of parameter sets ($\boldsymbol{\theta}_i$), we are able to cast them into one(two)-dimensional histograms to measure single-parameter marginal posterior distributions or two-parameter correlations.

\begin{figure*}[!hbtp]
    \centering
    \includegraphics[width=.99\textwidth]{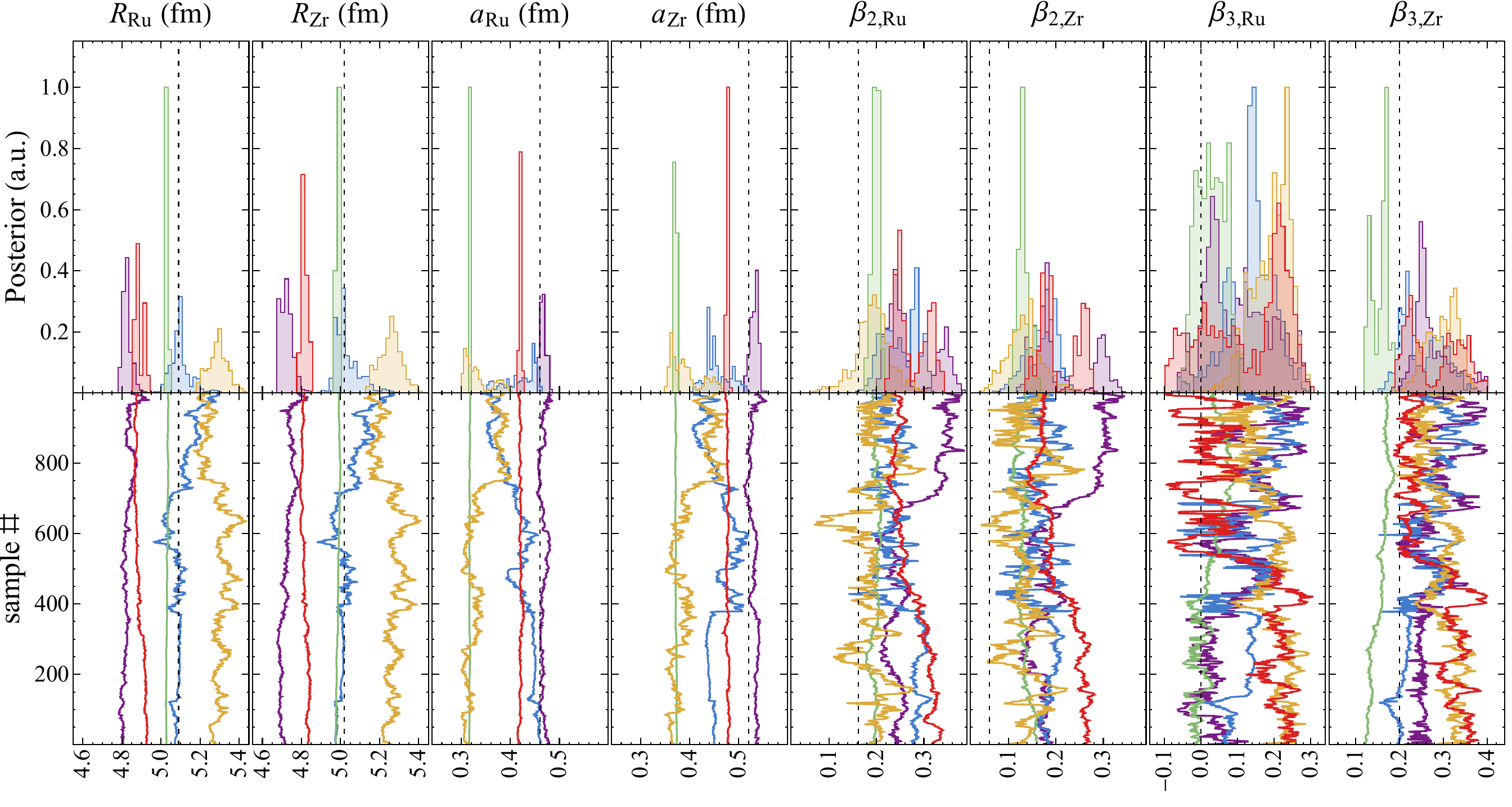}
    \caption{Marginal posterior distributions (top) and simulated trajectories (bottom) of nuclear structure parameters reconstructed from the mock data of the RuRu-to-ZrZr ratios of multiplicity distribution and elliptic, triangular, and radial flows. See Eq.~\protect{\eqref{eq:Ratio1}}. Different colors represents results from different Markov chains.
    \label{fig:Ratio1}}
    \includegraphics[width=.99\textwidth]{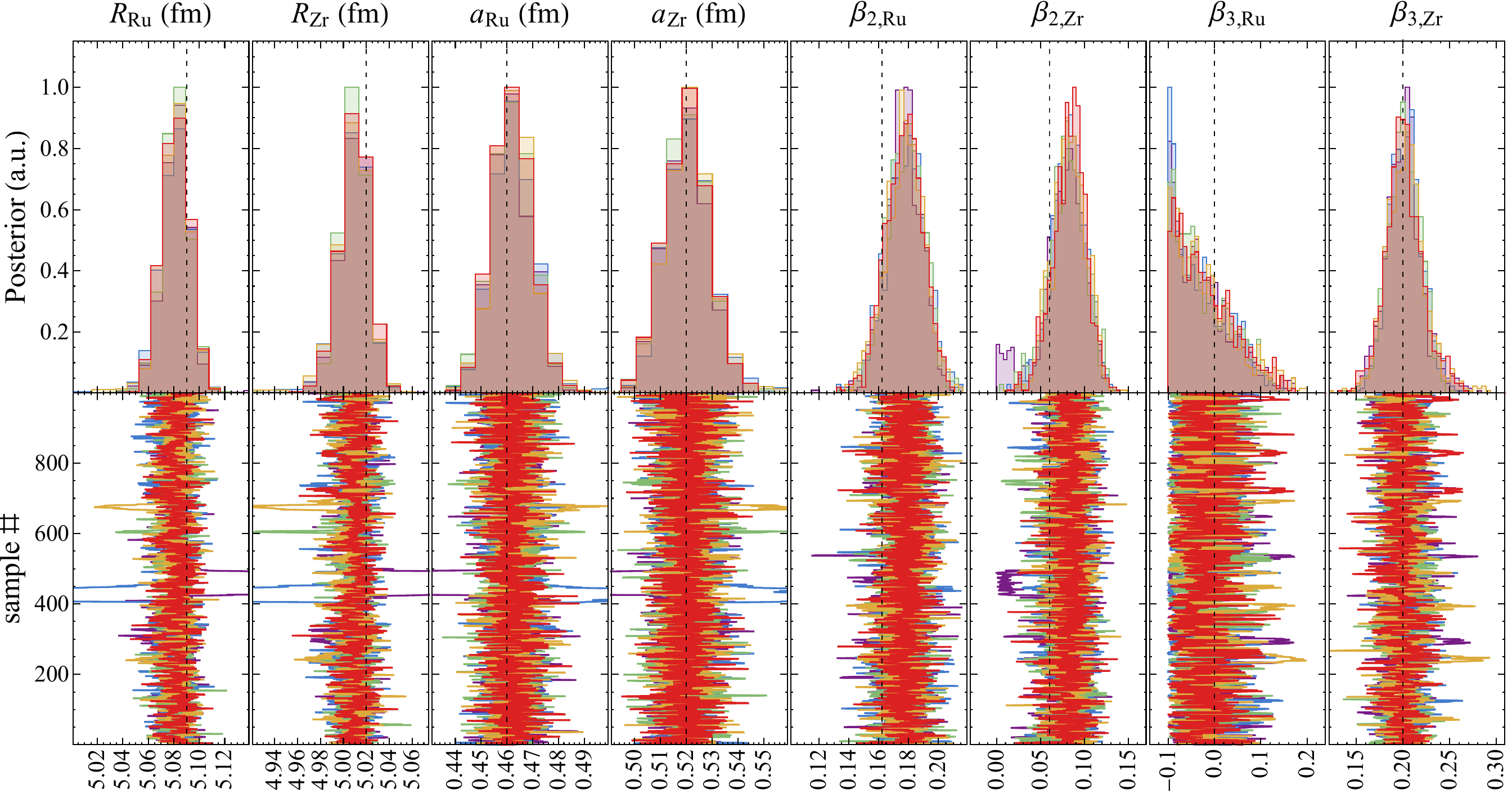}
    \caption{Same as Fig.~\protect{\ref{fig:Ratio1}} but reconstructed from single-system multiplicity distributions and the RuRu-to-ZrZr ratios of elliptic, triangular, and radial flows. See Eq.~\protect{\eqref{eq:Ratio2}}. 
    \label{fig:Ratio2}}
\end{figure*}

\section{Results}\label{sec:result}
With the framework set up in above, now we are ready to study the two questions raised in the Introduction. In this section, we move on to discuss the feasibility of nuclear structure reconstruction from heavy-ion collision observables. We will also investigate what observables are essential to a successful reconstruction. 

Observables of interest include charge multiplicity distribution($P_a$), elliptic($\varepsilon_{2,a}$), triangular($\varepsilon_{3,a}$) and radial($d_{\perp,a}$) flows for single collision system and their ratios in isobaric systems (RuRu-to-ZrZr). Here the index $a \in \{1-10, \cdots, 391-400\}$ labels the multiplicity bin. We denote the RuRu-to-ZrZr ratio of elliptic flow as $R_{\varepsilon_2, a} \equiv \varepsilon_{2, a}^\text{Ru}\, \big/\, \varepsilon_{2, a}^\text{Zr}$. Similarly, ratios of multiplicity distribution, triangular flow, radial flow are represented as $R_{P, a}$, $R_{\varepsilon_3, a}$, and $R_{d_{\perp}, a}$, respectively. For any observable $X$ we use $\widetilde{X}$ to represent the emulator evaluation for it.

The mock ``experiment measurements'' in this work is generated from two high statistic ($10^8$ events) MC-Glauber simulations taking the optimal parameter sets of RuRu and ZrZr isobaric system from~\cite{Jia:2021oyt}, which are respectively $R_\text{Ru} = 5.09~\text{fm}$, $a_\text{Ru} = 0.46~\text{fm}$, $\beta_{2,\text{Ru}} = 0.162$, $\beta_{3,\text{Ru}} = 0$ and  $R_\text{Zr} = 5.02~\text{fm}$, $a_\text{Zr} = 0.52~\text{fm}$, $\beta_{2,\text{Zr}} = 0.06$, $\beta_{3,\text{Zr}} = 0.20$. We take the diagonal of the statistical covariance matrix as the ``measurement'' uncertainty. They are $\sim (1/10)$ of the black curves in Fig.~\ref{fig:difference}, which can be computed from the number of events. Meanwhile, the theorectical uncertainties are taken to be the emulator uncertainties. Their exact values depend on the input parameter, but they are of the same order of magnitude as the red curves in Fig.~\ref{fig:difference}.

\subsection{Single System Reconstruction}
Let us begin with single systems and try to answer the first question in the Introduction.
That is, we would like to infer $R$, $a$, $\beta_{2}$, $\beta_{2}$, and their uncertainties from single system observables $P_a, \varepsilon_{2,a}, \varepsilon_{3,a}, d_{\perp,a}$. 
The Posterior distribution is given by Eq.~\eqref{eq:posterior}, with evidences in
the $\chi^2$ function~\eqref{eq:chisq_mat} being
\begin{align}
\begin{split}
    \boldsymbol{y}_\text{Ru} \equiv \big\{P_{a}^\text{Ru}, \varepsilon_{2, a}^\text{Ru}, \varepsilon_{3, a}^\text{Ru}, d_{\perp,a}^\text{Ru} \big\}_{a=1,\cdots,40}\,,
\end{split}
\end{align}
for the Ru-Ru system and likewise for the Zr-Zr system.
We take uniform prior for the parameter range of interest.


\begin{figure*}[!hbtp]
    \centering
    \includegraphics[width=0.75\textwidth]{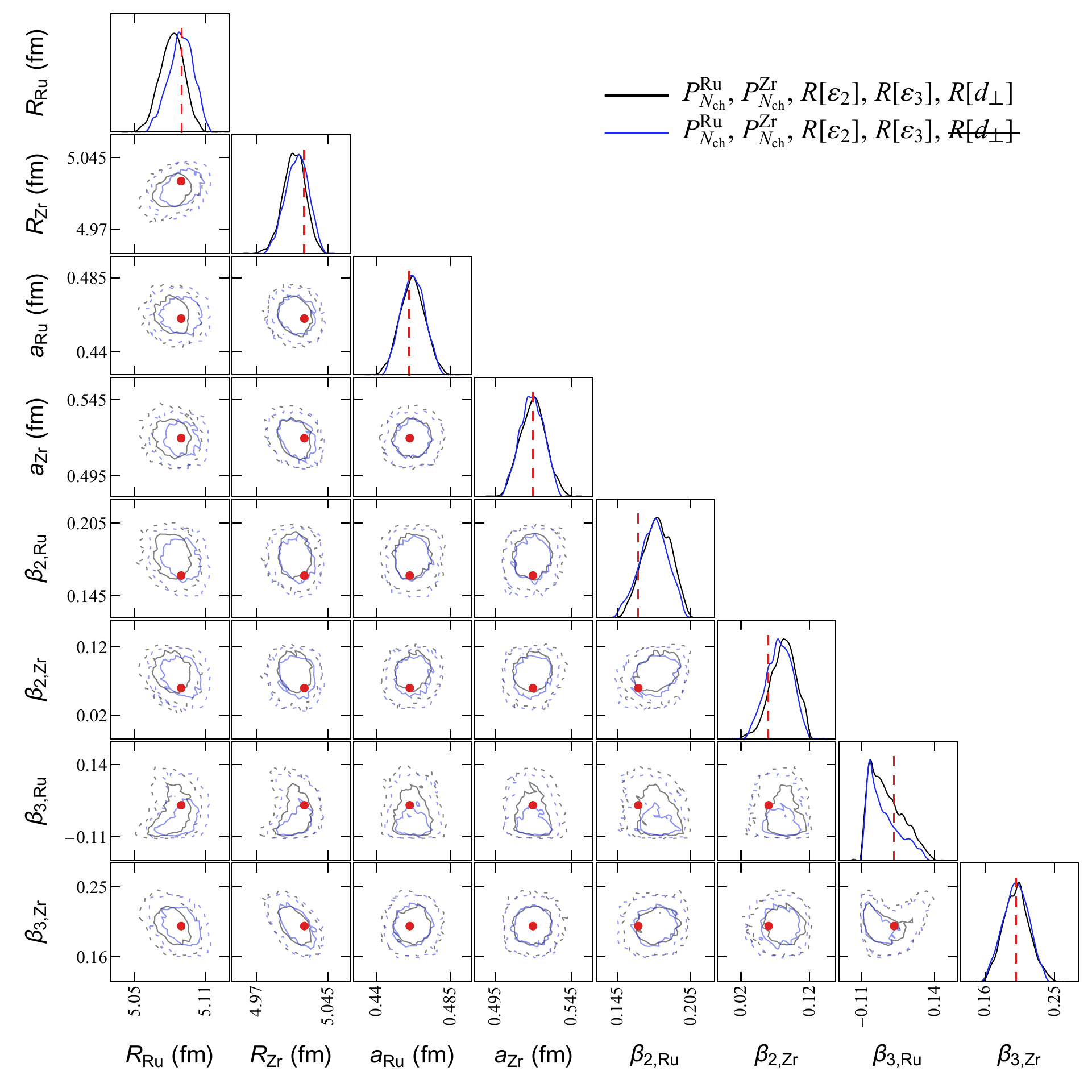}
    \caption{One- and two-dimensional marginal posterior distributions for reconstruction with[black, Eq.~\eqref{eq:Ratio2}] or without[blue, Eq.~\eqref{eq:Ratio3}] $d_\perp$-ratio. In the two-dimensional posterior, solid(dashed) curves correspond to the $68\%$($95\%$) C.L. contours.
    \label{fig:compare_dperp_Ratio}}
\end{figure*}

Our Bayesian Inference results of Ru and Zr Woods--Saxon parameters are shown in Fig.~\ref{fig:reconstruction_single}, each of which is obtained from MCMC simulation with five Markov chains of 1000 samples.
The marginal posterior distributions of each nuclear structure parameter shown as the smoothed histograms are plotted in the diagonal panels with the red dash vertical lines representing the ground truth of the parameter, while the posterior of two-parameter correlations are shown in the off-diagonal panels. From these two-dimension posterior distributions, we see that the nuclear structure parameters are not strongly correlated for the single system. Meanwhile, after sufficient iterations the posterior of the nuclear structure parameters converges to the stationary distribution which can be reflected from the one-dimension marginal posterior distributions. The mean values of
the posterior distribution for the parameters are $R_\text{Ru} = 5.09~\text{fm}$, $a_\text{Ru} = 0.4609~\text{fm}$, $\beta_{2, \text{Ru}} = 0.1599$, $\beta_{3, \text{Ru}} = 0.02304$ and $R_\text{Zr} = 5. 018~\text{fm}$, $a_\text{Zr} = 0.5217~\text{fm}$, $\beta_{2, \text{Zr}} = 0.05786$, $\beta_{3, \text{Zr}} = 0.1996$, which are close to the ground truth values we set. The ground truth of the parameters are also well covered by the packets of the Posterior distributions.

The inference performance on single systems in Fig.~\ref{fig:reconstruction_single} indicates that \textit{it is possible to reconstruct the nuclear structure from the final state observablses in heavy ion collisions,} given a reliable physical model to map the former to the latter. In other words, the mapping is reversible for the chosen observables and targeted nuclear structure parameters. However, it is well-known that the initial to final mapping relies on the details of the dynamical evolution of QGP. For instance, hydrodynamic simulation assuming different transport parameters may result in different final state observables. Therefore, reconstructions from single system observables are inevitably dependent on the specific model that describes or approximates the evolutionary dynamics. Nevertheless, it is claimed~\cite{Jia:2021oyt,Jia:2021qyu,Jia:2021tzt,Jia:2021wbq,Zhang:2021kxj,Zhang:2022fou} that the ratio of final state observables in an isobar system can largely cancel the dynamical-model dependence. 
So we move on to study what observables are needed to simultaneously extract the initial state nucleon distribution for a pair of isobar systems.

\subsection{Simultaneous Reconstruction for Isobar Systems}
Our attempts of simultaneously inferring nuclear structure for the two isobaric collision systems (Ru-Ru and Zr-Zr) start from their ratios on the four observables, i.e., evidences in the $\chi^2$ function~\eqref{eq:chisq_mat} read
\begin{align}
\begin{split}
    \boldsymbol{y}_\text{r,1} \equiv \big\{ R_{P,a}, R_{\varepsilon_2,a}, R_{\varepsilon_3,a},R_{d_{\perp},a} \big\}_{a=1,\cdots,40}\,.
\end{split}
\label{eq:Ratio1}
\end{align}
In the lower panels of Fig.~\ref{fig:Ratio1} we show the nuclear deformation parameters' trajectories of five Markov chains, represented by different colors and with the corresponding marginal posterior distribution in the upper panel of the figure. We note that the trajectories of different chains span a wide range in the parameter space and are well separated. Some of them do not overlap with each other, also we found their marginal posterior distributions do not necessarily cover the ground truth. These indicate that with the chosen evidence (i.e., ratio of those 4 observables) the MCMC can not converge to a stationary distribution of the Woods--Saxon parameters. 

We also observe strong correlations for the same parameters in the two isobar systems. For instance, the trajectories for $R_\text{Ru}$ and $R_\text{Zr}$ in the same Markov chain are with concurrent and highly similar behavior, and similar phenomenon also exhibit for other nuclear structure parameters. Such correlations indicate the degeneracy for the inference results.
The ratios of observables in isobar systems are mostly sensitive to the relative difference in their nuclear structure. One can change the parameters of the two isobar systems with the same amount and still reproduce the ratios of observables reasonably. \textit{Consequently, starting from purely the ratios, one can not simultaneously determine the nuclear structures of the two isobar systems.}

We note that the final state multiplicity can be well determined by the initial state total energy, which is also not sensitive to the transport parameters in the QGP evolution. Hence, we propose to include as well single-system charge multiplicity distributions besides other ratios for the purpose of model-independent and simultaneous inference of isobar nuclear structures.
If we reconstruct nuclear structure parameters based on charged multiplicity distributions for both RuRu and ZrZr systems, together with the RuRu-to-ZrZr ratios  of $\varepsilon_2$, $\varepsilon_3$, and $ d_{\perp}$, i.e., evidences in the $\chi^2$ function~\eqref{eq:chisq_mat} are chosen to be
\begin{align}
\begin{split}
    \boldsymbol{y}_\text{r,2} \equiv \big\{ P_{a}^\text{Ru}, P_{a}^\text{Zr},  R_{\varepsilon_2,a}, R_{\varepsilon_3,a},R_{d_{\perp},a} \big\}_{a=1,\cdots,40}\,,
\end{split}
\label{eq:Ratio2}
\end{align}
we find good convergence of Bayesian Inference results, as shown in Fig.~\ref{fig:Ratio2} (bottom) for the sampled trajectories. The marginal posterior distribution of each parameter spans a relatively narrow range, and all of them cover the ground truth as seen from Fig.~\ref{fig:Ratio2} (top). Especially, the parameter range covered by the trajectories are much narrower compared to Fig.~\ref{fig:Ratio1}. Therefore, \textit{taking the multiplicity distributions of the two isobar systems together with the ratios of $\varepsilon_2, \varepsilon_3$, and $d_{\perp}$, as shown in Eq.~\eqref{eq:Ratio2},
one can infer the isobar nuclear structures to very high precision.}
Meanwhile, strong correlations between the same parameter are still observed in the trajectories and more quantitatively in the two-dimensional marginal posterior distribution as shown as black curves in Fig.~\ref{fig:compare_dperp_Ratio}.

To test the robustness, we drop the ratio of $d_{\perp}$ in the Bayesian Inference, i.e., we take
\begin{align}
\begin{split}
    \boldsymbol{y}_\text{r,3} \equiv \big\{ P_{a}^\text{Ru}, P_{a}^\text{Zr},  R_{\varepsilon_2,a}, R_{\varepsilon_3,a} \big\}_{a=1,\cdots,40}
\end{split}
\label{eq:Ratio3}
\end{align}
as the evidences in the $\chi^2$ function~\eqref{eq:chisq_mat}.
We note that when dropping $d_{\perp}$-ratio in the evidence, the PC analysis needs to be performed separately since $d_{\perp}$ observables needs to be excluded therein. The corresponding one- and two-dimensional marginal posterior distributions are shown as blue curves in Fig.~\ref{fig:compare_dperp_Ratio}, which are very similar to the black ones. Particularly, the one-dimensional distributions almost overlap with each other for the two inference results, indicating that the Bayesian Inference based on \eqref{eq:Ratio2} and \eqref{eq:Ratio3} lead to consistent mean values and uncertainties of the Woods-Saxon parameters.
We conclude that \textit{the radial flow ($\langle p_T\rangle$) in heavy ion collisions, which can be estimated by $d_{\perp}$, carries redundant information as the ratios of elliptic/triangular flows}, and it is non-essential for the reconstruction of nuclear structure given multiplicity and anisotropic flow measurements provided.

\section{Summary}\label{sec:summary}
Most atomic nuclei present intrinsic deformed shapes, characterized notably by quadrupole and octupole moments, and the deformation is hard to be precisely measured in low-energy nuclei experiments. 
The high-statistics isobar collision experiment performed by the STAR collaboration~\cite{STAR:2021mii} shows that the ratio of observables in the isobar system is very sensitive to the difference in their nuclear structures, which attracts interest in studying nuclear structure in isobar collisions.

In this work, we investigated the plausibility of reconstructing the nuclear structure, in a statistical sense, from the heavy-ion collision observables. 
We first design a workflow of Bayesian Inference of Woods--Saxon parameters from different observables. As an exploratory first-step analysis, we take the Monte Carlo Glauber model as the estimator to map the initial nucleon distribution to the final state observables. For single systems, we find that one can precisely infer Woods--Saxon parameters from the final observables ($P,\varepsilon_2,\varepsilon_3,d_{\perp}$), assuming that a reliable model is provided to map the former to the latter. Whereas for the isobar systems, we find that one can not simultaneously determine the nuclear structures only from the ratio of those observables. However, a high-precision reconstruction becomes possible if the single-system multiplicity distributions are provided. Meanwhile, the ratio of radial flow is found to be non-essential for the reconstruction. 
Such a feasibility analysis paves the way to applying models with dynamical evolution to infer nuclear structure from real experimental data. Our efforts with AMPT model are in progress and will be reported elsewhere.

As a side product, we also find that the Woods--Saxon parameter dependence in the final state observables can be well approximated by a quadratic \textit{regression}, within a relatively wide range of parameters. Such a finding simplifies the choice of parameter sets in future studies.

\vspace{10mm}
\textbf{Acknowledgment. } The authors thank Dr. Jiangyong Jia, Peilian Li, Jinfeng Liao, Wilke van der Schee, and Chunjian Zhang for useful discussions. The work is supported by (i) the China Scholarship Council (YC), (ii) the BMBF under the ErUM-Data project (KZ), (iii) the AI grant of SAMSON AG, Frankfurt (KZ), (iv) U.S. Department of Energy, Office of Science, Office of Nuclear Physics, grant No. DE-FG88ER40388 (SS), (v)  the National Natural Science Foundation of China under contract Nos. 11890710 and 12147101, (vi) the Walter Greiner Gesellschaft zur F\"orderung der physikalischen Grundlagenforschung e.V. through the Judah M. Eisenberg Laureatus Chair at Goethe Universit\"at Frankfurt am Main (HS), and (vii) Tsinghua University under grant no. 53330500923 (SS).
We also thank the donation of NVIDIA GPUs from NVIDIA Corporation.

\begin{appendix}
\section{Monte Carlo Glauber Modeling}\label{sec:mcglauber}
The MC-Glauber simulation package used in this work follows the same setting as the date validated iEbE-VISHNew hydro package~\cite{Shen:2014vra}, with the details listed as follows.
For minimal-bias events, 
\begin{itemize}
\item 1) Sample nucleons according to the deformed Woods--Saxon distribution~\eqref{eq:woodssaxon}\footnote{It shall be worth noting that in some practices, a minimum distance between any two nucleons is required. The minimum distance is set to zero in this work.}.

\item 2) Sample the impact parameter $b$. Then the center of the projectile is translated to $(x_P,y_P) = (-b/2,0)$, while that of the target to $(x_T,y_T) = (b/2,0)$.

\item 3) For any pair of projectile nucleon (located at $\mathbf{x}_{P,i}$) and target nucleon (located at $\mathbf{x}_{T,j}$), determine whether inelastic collision can happen according to probability $P(d_{ij}) = 1-\exp\big[-\frac{2\sigma_{eff}}{\sigma_{NN}} e^{- 2\pi d_{ij}^2 / \sigma_{NN}}\big]$, where $d_{ij} \equiv \sqrt{(x_{P,i}-x_{T,j})^2 + (y_{P,i}-y_{T,j})^2}$ is the transverse distance. 
We collect every nucleon-nucleon collision event into the set of \{binary collisions\}, located at $\mathbf{x}_{ij} \equiv (\mathbf{x}_{P,i} + \mathbf{x}_{T,j})/2$.
Meanwhile, for every nucleon evolved in binary collisions, we collect it into the set of \{participants\}.
Events without initial binary collision will be discarded. 

\item 4) Compute binary collision density for $i\in$\{binary collisions\}
\begin{equation}
\rho_\mathrm{bin}(\mathbf{x}) \equiv \sum_{i} \frac{w_{\mathrm{bin},i}}{2\pi w_N^2}  \exp[-{(\mathbf{x}-\mathbf{x}_i)^2}/{(2w_N^2)}].
\end{equation}

\item 5) Compute participant density for $i\in$\{participants\}
\begin{equation}
\rho_\mathrm{part}(\mathbf{x}) \equiv \sum_{i} \frac{w_{\mathrm{part},i}}{2\pi w_N^2} \exp[-{(\mathbf{x}-\mathbf{x}_{i})^2}/{(2w_N^2)}].
\end{equation}

\item 6) Compute energy density as the superposition of the former two
\begin{equation}
e(\mathbf{x}) = \frac{S_e}{\tau_\mathrm{ini}} \Big(
	\frac{1-\alpha_\mathrm{glb}}{2}\cdot\rho_\mathrm{part}(\mathbf{x})
	+ \alpha_\mathrm{glb}\cdot\rho_\mathrm{bin}(\mathbf{x}) \Big).
\end{equation}
\end{itemize}
In above equations, $w_N = 0.5$~fm is nucleon width, 
$\sigma_{NN}$ is the inelastic nucleon-nucleon cross-section, 
$\sigma_{eff}$ is the effective cross-section with finite nucleon width implemented, which is determined by ensuring 
\begin{equation}
\sigma_{NN} = 
	2\pi \int b\,\mathrm{d}b \,
	 \Big( 1-\exp\big(-\frac{2\sigma_{eff}}{ \sigma_{NN}} e^{- 2\pi b^2 / \sigma_{NN}}\big) \Big) \,.
\end{equation}
Besides, $w_{\mathrm{bin},i}$ and $w_{\mathrm{part},i}$ are fluctuating weights, allowing extra fluctuation on multiplicity, they follow the gamma distribution with expectation value of unity and shape parameters $k_\mathrm{bin} \equiv k_\gamma \cdot \alpha_\mathrm{glb}$ and $k_\mathrm{part} \equiv k_\gamma \cdot (1-\alpha_\mathrm{glb})/2$, respectively,
\begin{eqnarray}
w_{\mathrm{bin}} &\sim& P(w_\mathrm{bin}) = \frac{w^{k_\mathrm{bin}-1} \exp(- w / k_\mathrm{bin})} {\Gamma(k_\mathrm{bin})\,
	k_\mathrm{bin}^{k_\mathrm{bin}}} \,,\\
w_{\mathrm{part}} &\sim& P(w_\mathrm{part}) = \frac{w^{k_\mathrm{part}-1} \exp(- w / k_\mathrm{part})} {\Gamma(k_\mathrm{part})\,
	k_\mathrm{part}^{k_\mathrm{part}}}  \,.
\end{eqnarray}
We use parameters as follows: 
initial time $\tau_\mathrm{ini}=0.4$~fm,
the overall scaling factor $S_e\times(N_\text{ch}/E)=9.34$, 
mixing parameter $\alpha_\mathrm{glb} = 0.123$, 
and gamma scaling $k_\gamma = 1.275$ which are fitted according to the multiplicity distributions of the isobar experiment~\cite{STAR:2021mii}.

\section{Correlation between Ratios}\label{sec:correlation}

In the main text, we express the observables using principal components,
\begin{align}
O_{a}(\bth)
=\;&
    \mu_a + \sigma_a\, \sum_{f=1}^{\Npc} V_{af} \mathrm{PC}_{f}(\bth)\, , 
\end{align}
which can be rewritten in a compact vectorized form,
\begin{align}
\begin{split}
\boldsymbol{O}_{\bth}
=\;&
    \boldsymbol{\mu} + \mathbf{\Lambda}_\sigma \cdot \mathbf{V} \cdot \boldsymbol{PC}_{\bth}\, , 
\end{split}
\end{align}
where we use italic bold font for vectors, and roman bold face for matrices, and $\cdot$ represents the inner product between matrices and/or vectors. The parameter dependence has been denoted by subscript/superscript. Also, $\mathbf{\Lambda}$ represents diagonal matrices.
Emulator uncertainty between different principal components are uncorrelated, 
\begin{align}\begin{split} 
&\mathrm{Cov}[\mathrm{PC}_{f}(\bth), \mathrm{PC}_{f'}(\bth') ]\\
\equiv\;&    \langle \mathrm{PC}_{f}(\bth) \mathrm{PC}_{f'}(\bth') \rangle -
    \langle \mathrm{PC}_{f}(\bth) \rangle \langle \mathrm{PC}_{f'}(\bth') \rangle \\
=\;&   \delta_{ff'} C_f(\bth,\bth')
\end{split}\end{align}
where $\langle \cdot \rangle$ denotes statistical ensemble average.
$C_f(\bth,\bth')$ is the covariance function of $\mathrm{PC}_{f}$ at two different parameter points, $\bth$ and $\bth'$. It can be computed as
\begin{align}
\begin{split}
    C_f (\bth,\bth') = k(\bth, \bth') - \sum^d_{i,j=1} k(\bth, \bth_i)(K^{-1})_{i,j} k(\bth_j, \bth').
\end{split}
\end{align}

The covariance between different observables follows
\begin{align}
\begin{split}
    \mathrm{Cov}[O_{a}(\bth), O_{b}(\bth')] = \sigma_a \sigma_b \sum_f V_{af} V_{bf} C_f(\bth, \bth').
\end{split}
\label{eq:CovMat}
\end{align}

For single system ($\bth = \bth'$), we denote $\Sigma_{\bth,ab} \equiv \mathrm{Cov}[O_{a}(\bth), O_{b}(\bth)]$, and the $(f,f)$-th element of diagonal matrix $\mathbf{\Lambda}_{C,\bth}$ is $C_f(\bth, \bth)$, i.e.,
\begin{align}
    \mathbf{\Sigma}_{\bth} = \mathbf{\Lambda}_\sigma \cdot \mathbf{V} \cdot
    \mathbf{\Lambda}_{C,\bth} \cdot \mathbf{V}^T \cdot \mathbf{\Lambda}_\sigma\,.
\end{align}
The $\chi^2$ function is then
\begin{align}
\begin{split}
&   (\boldsymbol{O}_{\bth} - \boldsymbol{E})^T \cdot \mathbf{\Sigma}_{\bth}^{-1} \cdot (\boldsymbol{O}_{\bth} - \boldsymbol{E})
\\=\,&
    (\Delta\boldsymbol{\mu}^T + \boldsymbol{PC}_{\bth}^T\cdot \mathbf{V}^T \cdot \mathbf{\Lambda}_\sigma) \cdot\mathbf{\Sigma}_{\bth}^{-1}\cdot (\Delta\boldsymbol{\mu} + \mathbf{\Lambda}_\sigma \cdot \mathbf{V} \cdot \boldsymbol{PC}_{\bth})
\\=\,&
    (\Delta\boldsymbol{\mu}^T \cdot\mathbf{\Lambda}_\sigma^{-1} \cdot \mathbf{V} + \boldsymbol{PC}_{\bth}^T) \cdot \mathbf{\Lambda}_{C,\bth}^{-1} \cdot 
    (\mathbf{V}^T \cdot \mathbf{\Lambda}_\sigma^{-1} \cdot\Delta\boldsymbol{\mu} + \boldsymbol{PC}_{\bth})
\\=\,&
    (\boldsymbol{PC}_{\bth}^T - \boldsymbol{PCE}^T) \cdot \mathbf{\Lambda}_{C,\bth}^{-1} \cdot 
    (\boldsymbol{PC}_{\bth} - \boldsymbol{PCE}),
\end{split}
\end{align}
where $\boldsymbol{E}$ represents experimental data on single system, $\Delta\boldsymbol{\mu} \equiv \boldsymbol{\mu} - \boldsymbol{E}$, and $\boldsymbol{PCE} \equiv \mathbf{V}^T \cdot \mathbf{\Lambda}_\sigma^{-1} \cdot (\boldsymbol{E} - \boldsymbol{\mu}) $ is the PC transformation on experimental observables. Therefore, the $\chi^2$ function in single system is equivalent to direct comparison of PC's.

When comparing the isobar systems, the $\chi^2$ function becomes 
\begin{align}
\begin{split}
&   (\boldsymbol{R}^{\bth,\bth'} - \boldsymbol{RE})^T  \cdot \mathbf{\Sigma}_{\bth,\bth'}^{-1} \cdot (\boldsymbol{R}^{\bth,\bth'} - \boldsymbol{RE}),
\end{split}
\end{align}
where $RE_a \equiv E^{\rm Ru}_a / E^{\rm Zr}_a$ is the experimental ratio, and ${R}^{\bth,\bth'}_{a} \equiv {O}^{\bth}_{a} / {O}^{\bth'}_{a}$.
We define the covariance of two ratios as
\begin{align}
\begin{split}
 {\Sigma}^{\bth,\bth'}_{R;a,b} =  \mathrm{Cov}\Big[\frac{O_{a}(\bth)}{O_{a}(\bth')}, \frac{O_{b}(\bth)}{O_{b}(\bth')}\Big],
\end{split}
\end{align}
and the covariance between an observable and a ratio as
\begin{align}
 {\Sigma}^{\bth;\bth,\bth'}_{X;a,b} =  \mathrm{Cov}\Big[O_{a}(\bth), \frac{O_{b}(\bth)}{O_{b}(\bth')}\Big],
\end{align}
and
\begin{align}
 {\Sigma}^{\bth';\bth,\bth'}_{X;a,b} =  \mathrm{Cov}\Big[O_{a}(\bth'), \frac{O_{b}(\bth)}{O_{b}(\bth')}\Big].
\end{align}

After straightforward but tedious calculations, we find
\begin{align}
\begin{split}
{\Sigma}^{\bth,\bth'}_{R;a,b} =\;& 
    \frac{C_{a,b}^{\bth,\bth} 
    - {R}^{\bth,\bth'}_{a} C_{a,b}^{\bth',\bth} 
    - {R}^{\bth,\bth'}_{b} C_{a,b}^{\bth,\bth'}
    + {R}^{\bth,\bth'}_{a} {R}^{\bth,\bth'}_{b} C_{a,b}^{\bth',\bth'}}{O_{a}^{\bth'} O_{b}^{\bth'}}
\,,
\end{split}
\\
    {\Sigma}^{\bth; \bth,\bth'}_{X;a,b} 
=\;&  \frac{ C_{a,b}^{\bth,\bth}}{O_{b}(\bth')}
-  \frac{R_{b}^{\bth,\bth'} C_{a,b}^{\bth,\bth'}}{O_{b}(\bth')},
\\
    {\Sigma}^{\bth'; \bth,\bth'}_{X;a,b} 
=\;&  \frac{ C_{a,b}^{\bth',\bth}}{O_{b}(\bth')}
-  \frac{R_{b}^{\bth,\bth'} C_{a,b}^{\bth',\bth'}}{O_{b}(\bth')},
\end{align}
where we have used the short-hand that $C_{a,b}^{\bth,\bth'} \equiv \mathrm{Cov}[O_{a}(\bth), O_{b}(\bth')]$ as defined in~\eqref{eq:CovMat}.


\end{appendix}

\bibliography{main}

\end{document}